\documentclass{sigchi}

\CopyrightYear{2018}
\toappear{}

\usepackage{balance}       % to better equalize the last page
\usepackage{graphics}      % for EPS, load graphicx instead 
\usepackage[T1]{fontenc}   % for umlauts and other diaeresis
\usepackage{txfonts}
\usepackage{mathptmx}
\usepackage[pdflang={en-US},pdftex]{hyperref}
\usepackage{color}
\usepackage{booktabs}
\usepackage{textcomp}

\usepackage{microtype}        % Improved Tracking and Kerning
\usepackage{ccicons}          % Cite your images correctly!
\usepackage[utf8]{inputenc} % for a UTF8 editor only

\usepackage{todonotes}
\usepackage{listings}
\def\plaintitle{A Survey on Deep Learning Toolkits and Libraries \\for Intelligent User Interfaces}
\def\plainauthor{Jan Zacharias, Michael Barz, Daniel Sonntag}
\def\plainkeywords{Artificial intelligence; Machine learning; Deep
learning; Interactive machine learning; Hyper-parameter tuning; Convolutional neural networks}
\makeatletter
\def\url@leostyle{%
  \@ifundefined{selectfont}{
    \def\UrlFont{\sf}
  }{
    \def\UrlFont{\small\bf\ttfamily}
  }}
\makeatother
\def\pprw{8.5in}
\def\pprh{11in}

\definecolor{linkColor}{RGB}{6,125,233}
\hypersetup{
  pdftitle={\plaintitle},
  pdfauthor={\plainauthor},
  pdfkeywords={\plainkeywords},
  pdfdisplaydoctitle=true,
  bookmarksnumbered,
  pdfstartview={FitH},
  colorlinks,
  citecolor=black,
  filecolor=black,
  linkcolor=black,
  urlcolor=linkColor,
  breaklinks=true,
  hypertexnames=false
}

\begin{document}

\title{\plaintitle}

\numberofauthors{3}
\author{%
  \alignauthor{Jan Zacharias\\
    \affaddr{German Research Center for Artificial Intelligence (DFKI)}\\
    \affaddr{Saarbrücken, Germany}\\
    \email{jan.zacharias@dfki.de}}\\
  \alignauthor{Michael Barz\\
    \affaddr{German Research Center for Artificial Intelligence (DFKI)}\\
    \affaddr{Saarbrücken, Germany}\\
    \email{michael.barz@dfki.de}}\\
  \alignauthor{Daniel Sonntag\\
    \affaddr{German Research Center for Artificial Intelligence (DFKI)}\\
    \affaddr{Saarbrücken, Germany}\\
    \email{daniel.sonntag@dfki.de}}\\
}

\maketitle

\begin{abstract}
This paper provides an overview of prominent deep learning toolkits and, in particular, reports on recent publications that contributed open source software for implementing tasks that are common in intelligent user interfaces (IUI).
We provide a scientific reference for researchers and software engineers who plan to utilise deep learning techniques within their IUI research and development projects.
\end{abstract}

{}


\section{Introduction}
Intelligent user interfaces (IUIs) aim to incorporate intelligent automated capabilities in human-computer interaction (HCI), where the net impact is an interaction that improves performance or usability in critical ways. Deep learning techniques can be used in an IUI to implement artificial intelligence (AI) components that effectively leverage human skills and capabilities, so that human performance with an application excels~\cite{DBLP:journals/corr/Sonntag17}.

Many IUIs, especially in the smartphone domain, use multiple input and output modalities for more efficient, flexible and robust user interaction \cite{2017:IST:3015783.3015785}. They allow users to select a suitable input mode, or to shift among modalities as needed during the changing physical contexts and demands of continuous mobile use. 

Deep learning has the potential to increase this flexibility with user and adaptation models that support speech, pen, (multi-) touch, gestures and gaze as modalities and can learn the appropriate alignment of them for mutual disambiguation. This ensures a higher precision in understanding the user input and to overcome the limitations of individual signal or interaction modalities \cite{Ngiam:2011}.

Deep learning systems are able to process very complex real-world input data by using a nested hierarchy of concepts with increasing complexity for its representation~\cite{Goodfellow-et-al-2016}.
Multimodal IUIs can greatly benefit from deep learning techniques because the complexity inherent in high-level event detection and multimodal signal processing can be modelled by likewise complex deep network structures and efficiently trained with nowadays available GPU-infrastructures.
Especially recurrent model architectures are suitable for processing sequential multimodal signal streams, e.g., for natural dialogues and long-term autonomous agents \cite{Sonntag09}.

\begin{figure*}
	\centering
	\includegraphics[width=1\linewidth]{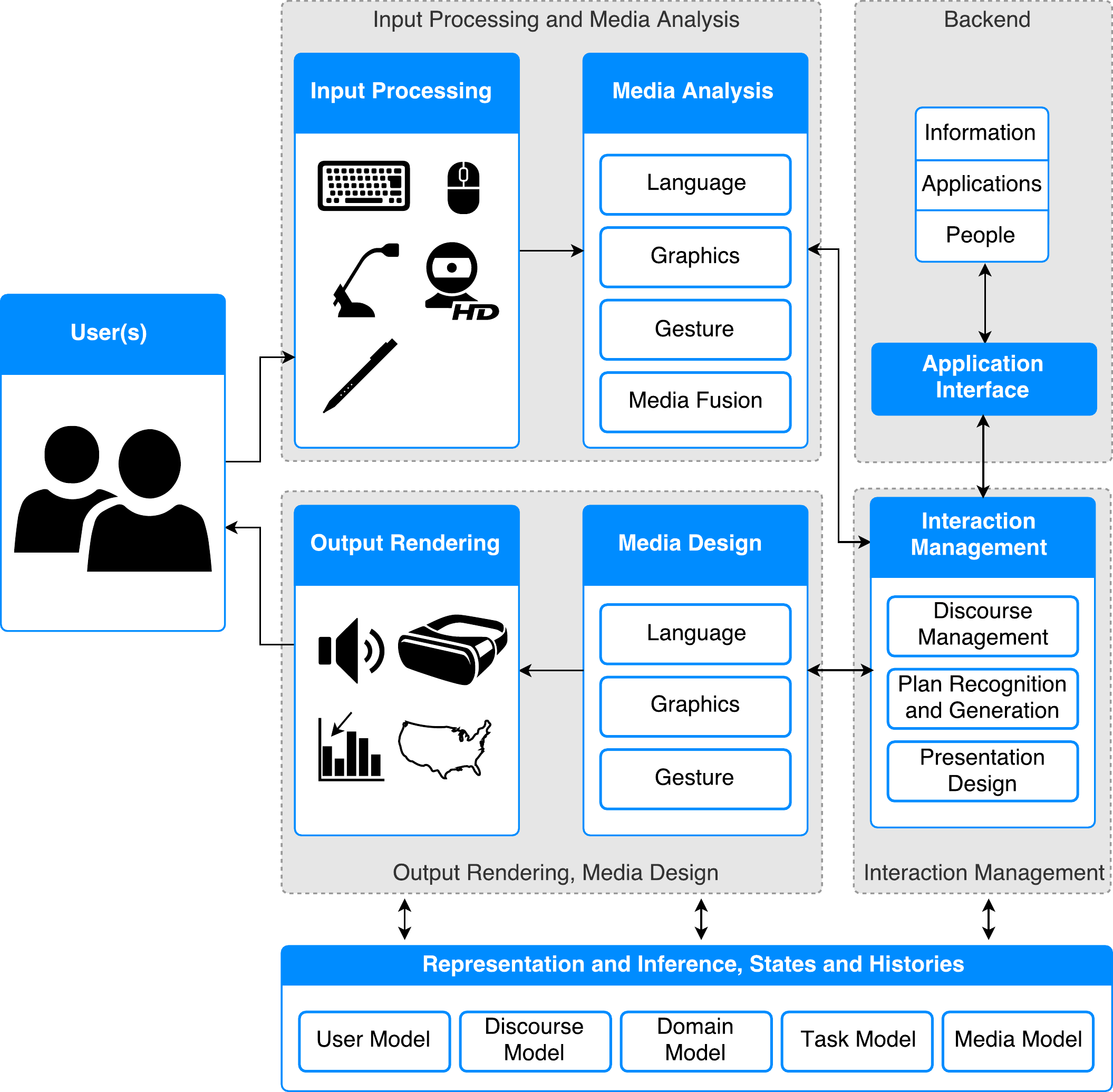}
	\caption{Categorization of intelligent user interface key components, based on the conceptual architecture \protect\cite{Maybury1998} and DFKI's Smartweb system \protect\cite{SonntagEHPPRR07}.}	
	\label{fig:tasks}
\end{figure*}

This paper provides IUI researchers and practitioners with an overview of deep learning toolkits and libraries that are available under an open source license and describes how they can be applied for intelligent multimodal interaction (considering the modules of the architecture in figure~\ref{fig:tasks} for classification).
Related deep learning surveys are in the medical domain\cite{Erickson2017} or focus on the techniques \cite{Parvat2017,Bahrampour2016,Shi2016}.
%WE focus okus auf IUI, mit relevanten Veröffentlichungen, auch IML Bezug

A major challenge common to all AI systems is to move from closed to open world settings.
Superhuman performance in one environment can lead to unexpected behaviour and dangerous situations in another: \begin{quote}>>The 'intelligence' of an artificial intelligence system \\ can be deep but narrow.<< \cite{CommitteeonTechnology2016}\end{quote}
An implication is the need for efficient learning methods and adaptive models for long-term autonomous systems.
Promising techniques include active learning~\cite{Settles2010}, reinforcement learning~\cite{NIPS2016_6420}, interactive machine learning~\cite{Amershi2014} and machine teaching~\cite{machine-teaching-new-paradigm-building-machine-learning-systems}.
We motivate the use of interactive training approaches enabling efficient and continuous updates to machine learning models.
This is particularly useful for enhancing user interaction because it enables robust processing of versatile signal streams and joint analysis of data from multiple modalities and sensors to understand complex user behaviour.

\section{Toolkits and Libraries}
We briefly introduce the most popular frameworks used for implementing deep learning algorithms (summarised in table~\ref{tab:comp}).
We include information about licenses and supported programming languages for quick compatibility or preference checks.
Then, we present and qualitatively evaluate open source contributions that are based on these frameworks and that implement key elements of an IUI.
Works are grouped into categories in alignment to an adapted version of the high-level IUI architecture as depicted in figure~\ref{fig:tasks}.
Other ways of applying the described systems are certainly possible and appropriate depending on the use case at hand.

\subsection{Deep Learning Frameworks}
The core of most deep learning contributions are sophisticated machine learning frameworks like TensorFlow~\cite{tensorflow2015-whitepaper} and Caffe~\cite{Jia2014}. 
Most of these software packages are published as open source, allowing independent review, verification, and democratised usage in IUI projects.
This does not imply that there are no errors in current implementations, however, the open character enables everybody to search for and eventually identify the root cause of wrong results or poor performance. 
Similarly, new features can be proposed and contributed. 
Table~\ref{tab:comp} lists toolkits and libraries that are a representative selection of available open source solutions. 
The table is sorted by a popularity rating proposed by François Chollet, the author of the Keras deep learning toolkit. GitHub metrics are used and weighted by coefficients such that the relative correlation of each metric reflects the number of users.
To model the factors, Chollet informally took into account Google Analytics data for the Keras project website, PyPI download data, ArXiv mentions data as well as Google Trends data among other sources. 
The rating is calculated as the sum of the GitHub Contributions$\times30$, Issues$\times20$, Forks$\times3$ and the Stars, scaled to $100\%$ defined by the top-scorer TensorFlow as a benchmark. 
While the exact numbers have been chosen manually, like in an ensemble model the relative order of magnitude of the coefficients matter beside the fact that multiple data sources are used.

\begin{table*}
	\label{tab:comp}
	\centering
	\renewcommand{\arraystretch}{1.5}
	\begin{small}
			\begin{tabular}{lp{3.5cm}p{2.4cm}llp{2cm}p{1cm}}
			\textbf{Name} & \textbf{Website}                                   & \textbf{GitHub URL} & \textbf{License} & \textbf{Language} & \textbf{APIs}              & \textbf{Rating [\%]} \\ \hline\hline
			TensorFlow \cite{tensorflow2015-whitepaper}      & \href{http://tensorflow.org}{http://tensorflow.org}                              &  \href{http://github.com/tensorflow/tensorflow}{tensorflow/\newline tensorflow}          & Apache-2.0         & C++, Python         & Python, C++, Java, Go              & 100                     \\ \hline
			Keras \cite{chollet2015keras}          & \href{http://keras.io/}{http://keras.io/ }                                                              & \href{http://github.com/fchollet/keras}{fchollet/keras}             & MIT                & Python              & Python, R                          & 46.1                    \\ \hline
			Caffe \cite{Jia2014}          & \href{http://caffe.berkeleyvision.org}{http://caffe.berkeleyvision.org}                                          & \href{http://github.com/BVLC/caffe}{BVLC/caffe}         & BSD                & C++                 & Python, MATLAB                     & 38.1                    \\ \hline
			MXNet \cite{Chen2015a}          & \href{http://mxnet.io}{http://mxnet.io}                                                               & \href{http://github.com/apache/incubator-mxnet}{apache/incubator-mxnet}                     & Apache-2.0         & C++                 & Python, Scala, R, JavaScript, Julia, MATLAB, Go, C++, Perl & 34                      \\ \hline
			Theano \cite{TheTheanoDevelopmentTeam2016}         & \href{http://deeplearning.net/software/theano}{http://deeplearning.net/\newline software/theano}                                       & \href{http://github.com/Theano/Theano}{Theano/Theano}                  & BSD        & Python              & Python                             & 19.3                    \\ \hline
			CNTK \cite{Yu2015}           & \href{https://docs.microsoft.com/en-us/cognitive-toolkit}{https://docs.microsoft.com/\newline en-us/cognitive-toolkit}                     & \href{http://github.com/Microsoft/CNTK}{Microsoft/CNTK}                 & MIT                & C++                 & Python, C++, C\#, Java             & 18.4                    \\ \hline
			DeepLearning4J \cite{Deeplearning4jDevelopmentTeam}  & \href{https://deeplearning4j.org}{https://deeplearning4j.org}                                                      & \href{http://github.com/deeplearning4j/deeplearning4j}{deeplearning4j/\newline deeplearning4j}  & Apache-2.0         & Java, Scala           & Java, Scala, Clojure, Kotlin       & 17.8                    \\ \hline
			PaddlePaddle    & \href{http://www.paddlepaddle.org}{http://www.paddlepaddle.org}                                                    & \href{http://github.com/baidu/paddle}{baidu/paddle}                   & Apache-2.0         & C++                 & C++                                & 16.3                    \\ \hline
			PyTorch         & \href{http://pytorch.org}{http://pytorch.org}                                                              & \href{http://github.com/pytorch/pytorch}{pytorch/pytorch}                & BSD        & C++, Python         & Python                             & 14.3                    \\ \hline
			Chainer \cite{Tokui2015}        & \href{https://chainer.org}{https://chainer.org}                                                             & \href{http://github.com/pfnet/chainer}{pfnet/chainer}                  & MIT                & Python              & Python                             & 7.9                     \\ \hline
			Torch7 \cite{Collobert2011}         & \href{http://torch.ch}{http://torch.ch/}                                                               & \href{http://github.com/torch/torch7}{torch/torch7}                   & BSD                & C, Lua              & C, Lua, LuaJIT                     & 7.8                     \\ \hline
			DIGITS \cite{Yeager2015}         & \href{https://developer.nvidia.com/digits}{https://developer.nvidia.com/\newline digits}                               & \href{http://github.com/NVIDIA/DIGITS}{NVIDIA/DIGITS}                  & BSD                & Python              & REST/Json                          & 7.8                     \\ \hline
			TFLearn \cite{tflearn2016}        & \href{http://tflearn.org}{http://tflearn.org}                                                              & \href{http://github.com/tflearn/tflearn}{tflearn/tflearn}                & MIT                & Python              & Python                             & 7.5                     \\ \hline
			Caffe2          & \href{https://caffe2.ai}{https://caffe2.ai}                                                               & \href{http://github.com/caffe2/caffe2}{caffe2/caffe2}                  & Apache-2.0         & C++, Python         & Python, C++                     & 7.4                     \\ \hline
			dlib \cite{King2009}           & \href{http://dlib.net}{http://dlib.net}                                                                 & \href{http://github.com/davisking/dlib}{davisking/dlib}                 & Boost              & C++                 & C++, Python                               & 5.7                    
		\end{tabular}
	\end{small}
	\caption{Open source software overview with rating based on GitHub metrics}
\end{table*}

\subsection{Deep Learning in IUIs}
One can find a multitude of deep learning software on the web and it is unclear whether these packages are useful and easy to setup in an IUI project at a first glance.
With this work, we want to shed light on selected open source contributions by providing an overview and sharing our experiences in terms of utility and ease of use.
We group related works based on the long acknowledged conceptual architecture by Wahlster and Maybury~\cite{Maybury1998} and its reference implementation \cite{SonntagEHPPRR07} which defines essential parts and modules of IUIs (see figure~\ref{fig:tasks}). 

We consider the functional coherent elements \emph{Input Processing \& Media Analysis}; \emph{Interaction Management, Output Rendering \& Media Design} and \emph{Backend} services as the main building blocks. 
Many works can be applied at different stages or implement multiple roles at once, particularly deep learning systems that are trained end-to-end.
Inference mechanisms and the representation of the current "information state" and histories of meta models are shared across all IUI components.

\subsubsection{Input Processing \& Media Analysis}
Components that implement this role help in analysing and understanding the user by including available modalities and fusing them, if appropriate.
Examples are IUIs that require a natural language understanding (NLU) component to extract structured semantic information from unstructured text, e.g., for classifying the user's intent and extracting entities. 

Hauswald et al. \cite{Hauswald2015} implemented the Tonic Suite that offers different classification services as a service based on DjiNN, an infrastructure for deep neural networks (DNN).
It can process visual stimuli for image classification based on AlexNet \cite{Krizhevsky2012} and face recognition by replicating Facebook's DeepFace \cite{Taigman2014}. 
Natural language input is supported in terms of digit recognition based on MNIST \cite{lecun1998gradient} and automated speech recognition (ASR) based on the Kaldi\footnote{https://github.com/kaldi-asr/kaldi} ASR toolkit \cite{Povey_ASRU2011} trained on the VoxForge\footnote{http://www.voxforge.org} open-source large scale speech corpora.
Further, several natural language processing techniques are available: part-of-speech tagging, named entity recognition and word chunking---all based on neural networks first introduced by NEC's SENNA\footnote{http://ml.nec-labs.com/senna} project. 
The Tonic suite relies on the Caffe framework. 

The BSD 3-Clause licensed C++ source code\footnote{https://github.com/claritylab/djinn} is however inconsistently annotated and TODO's to improve the documentation quality remain unresolved. Sufficient installation instructions for DjiNN and Tonic are separately hosted\footnote{http://djinn.clarity-lab.org/djinn/}$^,$\footnote{http://djinn.clarity-lab.org/tonic-suite/}. Both software packages have not been updated since 2015, rely on an outdated Caffe version and have many other legacy dependencies, thus extending installation time considerably.

Relation extraction is an important part of natural language processing (NLP): relational facts are extracted from plain text.
For instance, this task is required in the automated population and extension of a knowledge base from user input.
Lin et al.~\cite{Lin2016} released a neural relation extraction implementation\footnote{https://github.com/thunlp/NRE} under the MIT license. 
The same repository hosts the re-implementation of the relation extractors of Zeng et al. \cite{Zeng2014,Zeng2015}. 
The C++ code is completely undocumented, the ReadMe file lists comparative results whereas required resources are not directly accessible and dependencies are not listed.
More recently the authors published a better documented relation extractor \footnote{https://github.com/thunlp/TensorFlow-NRE} inspired by \cite{Lin2016,Zhou2016} which uses TensorFlow and is written in Python---the code is annotated and dependencies are listed appropriately.
Similarly, Nguyen and Grishman~\cite{Nguyen2015} proposed the combination of convolutional and recurrent neural networks in hybrid models via majority voting for relation extraction. 
The authors published their source code\footnote{https://github.com/anoperson/DeepIE}, which is based on the Theano framework, to allow others to verify and potentially improve on their method concerning the ACE 2005 Corpus\footnote{https://catalog.ldc.upenn.edu/ldc2006t06}.
The Python source does not contain licensing information and is not annotated. The ReadMe file outlines how the evaluation can be performed and states that the dataset needs to be procured separately, incurring a \$$4000$ fee.

Chen and Manning proposed a dependency parser using neural networks that analyses the grammatical structure of sentences and tries to establish relationships between "head" words and words which modify those heads \cite{Chen2014}. The software is part of the Stanford Parser\footnote{https://nlp.stanford.edu/software/lex-parser.html} and CoreNLP,\footnote{https://github.com/stanfordnlp/CoreNLP} a Java toolkit for NLP which allows the computer to analyse, understand, alter or generate natural language. Consequently, CoreNLP relates also to the media design element of an IUI. Building instructions are included in the ReadMe file and a well-written HTML documentation is available\footnote{https://stanfordnlp.github.io/CoreNLP/}.

The open source contribution RASA\_NLU\footnote{https://github.com/RasaHQ/rasa\_nlu}~\cite{Bocklisch2017}, written in Python and published under the Apache-2.0 license, performs natural language understanding with intent classification and entity extraction. Several machine learning backends can be employed: spaCy\footnote{https://github.com/explosion/spaCy} which uses thinc\footnote{https://github.com/explosion/thinc}, a deep learning capable library, MITIE\footnote{https://github.com/mit-nlp/MITIE} with dlib as backend and scikit-learn\footnote{https://github.com/scikit-learn/scikit-learn}. When using docker\footnote{https://www.docker.com/} the installation is noteworthy simple.
A single command downloads all required components and dependencies and starts the container with the NLU service in minutes (depending on the internet connection). A complete installation and getting started guide is available as ReadMe. As the source code is consistently annotated, an automatically generated Sphinx\footnote{http://www.sphinx-doc.org} HTML documentation is accessible\footnote{https://rasahq.github.io/rasa\_nlu/master/}. Short questions can be asked in a gitter chat\footnote{https://gitter.im/RasaHQ/rasa\_nlu}.

In pervasive computing, understanding how the user interacts with the environment is essential.
Bertasius et al.~\cite{Bertasius2016} designed a deep neural network model with Caffe\footnote{https://github.com/gberta/EgoNet} that processes the user's visual and tactile interactions for identifying the action object. The authors provide a pre-trained model and the Python code that predicts the area of the action object in an RGB(D) image, i.e., depth information is optional and can be used to improve the accuracy. Unfortunately, the annotated dataset that was created by the authors to train and test the model remains unpublished, thus preventing complete verification of the results and model enhancements by third parties. The published source is thoroughly annotated but lacks licensing information and detailed dependency information.
In a follow-up publication, Bertasius et al.~\cite{Bertasius2016a} demonstrated that the supervised creation of the training dataset can be omitted by using segmentation and recognition agents, implemented as cross-pathway architecture in a visual-spatial network (VSN). 
Unsupervised learning is desirable, because the training of the action-object detection model requires pixelwise annotation of captured images by humans, which is time-intensive and costly. The implemented VSN learns to detect likely action-objects from unlabelled egocentrically captured image data.  The GitHub repository\footnote{https://github.com/gberta/Visual-Spatial-Network} contains pre-trained models, the Python code to perform predictions and matlab sources for the multiscale combinatorial grouping that is required for the segmentation agent. Similarly to the first contribution, the code lacks a license while code annotations are sufficient. The ReadMe file contains general training instructions, the actual setup of the training toolchain is very cumbersome as merely pointers are given. 

In \cite{Barz2016} Barz and Sonntag used Caffe to combine gaze and egocentric camera data with GPU based object classification and attention detection for the construction of episodic memories of egocentric events in real-time. 
The code is not yet available, but will be published as a plug-in for the Pupil Labs eye tracking suite~\cite{Kassner:2014:POS:2638728.2641695}\footnote{https://github.com/pupil-labs/pupil} in mid of 2018.
%The code is available on github and documented like industrial code. 

\subsubsection{Interaction Management, Output Rendering \& Media Design}
User input is interpreted with regard to the current state of considered models, e.g., the discourse context, in order to identify and plan future actions and design appropriate IUI output. Components described here implement a central dialogue management functionality and generate outputs for the users in terms of multimodal natural language generation (NLG).

RASA\_CORE\footnote{https://github.com/RasaHQ/rasa\_core}~\cite{Bocklisch2017} is an open source discourse/dialogue manager framework which is written in Python and uses Keras' LSTM implementation in order to allow contextual, layered conversations. While no docker image is available, the documentation\footnote{https://core.rasa.ai/} quality is on par with RASA\_NLU, a chat is analogously available\footnote{https://gitter.im/RasaHQ/rasa\_core}.
Four example chatbots are provided\footnote{https://github.com/RasaHQ/rasa\_core/tree/master/examples} for getting started easily as well as a number of tutorials\footnote{https://core.rasa.ai/tutorial\_basics.html}. 
In an evaluation, Braun et al. \cite{Braun2017} found the RASA ensemble to score second best against commercial closed source systems. Note that RASA\_CORE also fits the media design module as it can generate responses and clarification questions in a conversational system on its own. The capabilities can be extended by using more sophisticated NLG techniques (see Gatt and Kramer~\cite{Gatt2018} for a recent overview).

In \cite{DeVries2016,Strub2017} a deep reinforcement learning system for optimising a visually grounded goal-directed dialogue system was implemented using TensorFlow. GuessWhat?!\footnote{https://github.com/GuessWhatGame/guesswhat/}, the corresponding open source contribution, is moderately annotated and contains a ReadMe file that allows verification of the published results by outlining required steps for their reproduction. Pre-trained models are available for download and basic installation instructions are contained as well. Unfortunately, the authors omitted correct Python dependency version information which leads to some trial-and-error during installation. This contribution uses the Apache-2.0 license.\footnote{Update in future revisions: https://github.com/voicy-ai/DialogStateTracking}

%User interfaces which are implemented as part of an augmented reality (AR) system in combination with a head mounted see-through display device, like Microsofts's HoloLens\footnote{https://www.microsoft.com/en-us/hololens}, require powerful spatial mapping capabilities in order to augment the reality in a geometrically correct way. 

\subsubsection{Backend}

Visual scene understanding is an important property of an IUI which needs to process image or video input. From a technical perspective, this requires image classification and accurate region proposals so that a meaningful segmentation of a scene with multiple objects is possible. Sonntag et al. contributed the py-faster-rcnn-ft\footnote{https://github.com/DFKI-Interactive-Machine-Learning/py-faster-rcnn-ft} Python library that allows convenient fine-tuning of deep learning models that offer this functionality. E.g., the VGG\_CNN\_M\_1024 model \cite{Chatfield2014} can be fine-tuned on specific categories of the MS COCO \cite{Lin2014} image dataset, hence improving classification accuracy \cite{Girshick2012}. While the library can work entirely in the background, it also comes with an user interface allowing to select categories and inspect the results graphically. The software uses the Caffe framework and is GPL-3 licensed. It features an extensive ReadMe file containing an installation and getting started guide with example listings and optional pre-trained model resources for quick evaluation.
This work is based on py-faster-rcnn\footnote{https://github.com/rbgirshick/py-faster-rcnn}~\cite{renNIPS15fasterrcnn}.

Nvidia's DIGITS\footnote{https://github.com/NVIDIA/DIGITS} enables deep learning beginners to design and train models to solve image classification problems and put these models to use. The Python software features an intelligent web (HTML/JS) interface that allows highly interactive modification of the neural network by the user as well as data management and fine-tuning of existing models. The DIGITS interface displays performance statistics in real time so that the user can quickly identify and select the best performing model for deployment. On the other hand, DIGITS can also be used as a backend component of an IUI via its REST API to perform inference on trained models. The software is BSD-3 licensed and can use Caffe, Torch, and Tensorflow as deep learning framework. The installation via docker requires little effort and the accompanying ReadMe file links to multiple well-grounded howto guides. A graphically enriched documentation and introduction to deep learning is available from Nvidia\footnote{https://devblogs.nvidia.com/parallelforall/digits-deep-learning-gpu-training-system/}.

\section{Outlook: Interactive Machine Learning}
To develop the positive aspects of artificial intelligence, manage its risks and challenges, and ensure that everyone has the opportunity to help in building an AI-enhanced society and to participate in its benefits, we suggest using a methodology where human intelligences \& machine learning take the center stage: {\it Interactive Machine Learning is the design and implementation of algorithms and intelligent user interface frameworks that facilitate machine learning with the help of human interaction.}

This field of research explores the possibilities of helping AI systems to achieve their full potential, based on interaction with humans. The current misconception is that artificial intelligence is more likely to be performance oriented than learning oriented. By machine teaching~\cite{machine-teaching-new-paradigm-building-machine-learning-systems} we can "assist" AI systems in becoming self-sustaining, "lifelong" learners \cite{Lieberman2001227,DBLP:journals/corr/Sonntag17} as a domain expert trains complex models by encapsulating the required mechanics of machine learning. This resource oriented methodology contributes in closing the gap between the demand for machine learning models and relatively low amount of experts capable of creating them.

Interactive machine learning (IML) includes feedback in real-time, allowing fast-paced iterative model improvements through intelligent interactions with a user~\cite{Amershi2014}.
As shown by Sonntag, when using artificial intelligence (AI) to implement intelligent automated capabilities in an IUI, effects on HCI must be considered in order to prevent negative side-effects like diminished predictability and lost controllability which ultimately impact the usability of the IUI \cite{Sonntag2012}. This adverse effect can be diminished by adopting the concept of the binocular view when building IUIs: both AI and HCI aspects are simultaneously addressed so that the systems intelligence, and how the user should be able to interact with it, are optimally synchronised \cite{Jameson2009}.

Many principles used in active learning are adopted, for example, query strategies for selecting most influential samples that shall be labelled~\cite{Fogarty2008,Settles2011} and semi-supervised learning for the automatic propagation of labels under confidence constraints~\cite{Settles2011}.
IML benefits from IUIs that support the user in training/teaching a model and is, at the same time, essential for building and maintaining models for intelligent and multimodal interaction.

Recent works use deep learning in combination with active or passive user input to improve the model training or performance:
Green et al. \cite{Green2015} applied the IML concept for a language translation task that benefits from human users and machine agents: the human in the loop can produce higher quality translations while the suggested machine translation is continuously improved as the human corrects its suggestions. 
Venkitasubramanian et al.~\cite{Venkitasubramanian2017} present a model that learns to recognise animals by watching documentaries. They implicitly involve humans by incorporating their gaze signal together with the subtitles of a movie as weak supervision signal. Their classifiers learns using image representations from pre-trained convolutional neural networks (CNN).
Jiang et al.~\cite{Jiang2017} present an algorithm for learning new object classes and corresponding relations in a human-robot dialogue using CNN-based features. The relations are used to reason about future scenarios where known faces and objects are recognised.
Cognolato et al.~\cite{Cognolato2017} use human gaze and hand movements to sample images of objects that get manipulated and fine-tune a CNN with that data.
In the context of active learning, Käding et al.~\cite{Kaeding2017} investigate the trade-off between model quality and the computational effort of fine-tuning for continuously changing models.
Jiang et al.~\cite{Jiang2017a} present a GPU-accelerated framework for interactive machine learning that allows easy model adoption and provides several result visualisations to support users.

In \cite{Jiang2017a} techniques for customised and interactive model optimisation are proposed. Jiang and Canny use the BIDMach framework \cite{Canny2013} which builds upon Caffe to provide a machine learning architecture which is modular and supports primary and secondary loss functions. The users of this system are able to directly manipulate deep learning model parameters during training. The user can perform model optimisations with the help of interactive visualisation tools and controls via a web interface. It however remains challenging to transfer the concepts of interactive machine learning to deep learning algorithms and to investigate their impact on model performance and usability, particularly in the context of IUIs and multimodal settings.

\bibliographystyle{SIGCHI-Reference-Format}
\bibliography{paper}
\end{document}